\documentclass[twocolumn,prl,showpacs,amsmath,amssymb,superscriptaddress, 10.5pt]{revtex4}

\usepackage{times}
\usepackage{graphicx}
\usepackage{dcolumn}
\usepackage{bm}
\usepackage{natbib}

\begin{document}

\title{Evidence for effective thermal boundary resistance from magnon/phonon disequilibrium}

\author{M.C. Langner}
\affiliation{Department of Physics, University of California, Berkeley, CA 94720}
\affiliation{Materials Science Division, Lawrence Berkeley National Laboratory, Berkeley, CA 94720}
\author{C.L.S. Kantner}
\affiliation{Department of Physics, University of California, Berkeley, CA 94720}
\affiliation{Materials Science Division, Lawrence Berkeley National Laboratory, Berkeley, CA 94720}
\author{Y.H. Chu}
\affiliation{Department of Materials Science and Engineering, University of California, Berkeley, CA 94720}
\author{L.M. Martin}
\affiliation{Materials Science Division, Lawrence Berkeley National Laboratory, Berkeley, CA 94720}
\author{P. Yu}
\affiliation{Department of Physics, University of California, Berkeley, CA 94720}
\author{R. Ramesh}
\affiliation{Department of Physics, University of California, Berkeley, CA 94720}
\affiliation{Department of Materials Science and Engineering, University of California, Berkeley, CA 94720}
\author{J. Orenstein}
\affiliation{Department of Physics, University of California, Berkeley, CA 94720}
\affiliation{Materials Science Division, Lawrence Berkeley National Laboratory, Berkeley, CA 94720}

\date{\today}

\begin{abstract}
We use the time-resolved magneto-optical Kerr effect (TRMOKE) to measure the local temperature and heat flow dynamics in ferromagnetic SrRuO$_{3}$ thin films. After heating by a pump pulse, the film temperature decays exponentially, indicating that the heat flow out of the film is limited by the film/substrate interface. We show that this behavior is consistent with an effective boundary resistance resulting from disequilibrium between the spin and phonon temperatures in the film.

\end{abstract}

\maketitle

Propagation of heat in magnetic nanostructures is a subject of importance for both fundamental and practical reasons. Regarding applications, devices based on spin-current torque dissipate significant amounts of energy in the process of flipping spins \cite{spintronics, racetrack}, and successful operation requires that energy escape from the structure in the form of heat. From a fundamental point of view, the factors that limit the flow of heat out of magnetic nanostructures are not well understood. For example, macroscopic (Fourier) modeling of heat flow is not applicable when sample dimensions are smaller than phonon or other quasiparticle mean-free-paths. Heat propagation is more complex in ferromagnetic nanostructures because thermal energy is shared over spin, electron, and phonon degrees of freedom (DOF).  The net rate of heat flow depends on the thermal diffusivity of each DOF, the rate at which energy is shared between them, and finally, the extent to which each DOF is confined to the nanostructure.

A deeper understanding of heat flow in magnetic nanostructures requires experiments that explore the dynamics of local temperature change over a wide range of parameters, for example, bath temperature and sample dimensions.  Optical techniques such as as thermal reflectometry have proven useful in measuring the local temperature in nanostructures. Dynamics are probed effectively in time-resolved versions of this technique, in which a focused laser pulse heats the sample and a second time-delayed pulse probes the temperature via the associated variation in optical reflectivity \cite{trr_hohlfeld, au_solution, nanoscalereview, trtransport}.  One drawback of time-resolved thermal reflectometry (TRR) is that it is limited to structures whose dimension, $L$, is greater than the penetration depth, $\lambda_p$, of the laser light.  If $L\ll\lambda_p$ most of the pump light is absorbed in surrounding materials rather than nanostructure, which leads to changes in reflectivity not associated with heating of the sample.

When the structures of interest are magnetic, the time-resolved magneto-optical Kerr effect (TRMOKE) offers potential advantages over TRR. This measurement uses the temperature dependence of the Kerr rotation, $\Delta \Theta_K$, rather than reflectivity, as the thermometer. The most significant advantage relative to TRR is that $\Delta \Theta_K$ arises only from the sample under study, even when $L\ll\lambda_p$.  A second advantage is that TRMOKE can be a more sensitive probe of temperature in magnets in which strong spin-orbit coupling gives rise to large Kerr effects.

In this work we use TRMOKE to study the dynamics of heat transport in ferromagnetic thin films of SrRuO$_{3}$ (SRO).  SRO is ideal for studying heat transport in nanoscale ferromagnets for several reasons. First, high-quality, epitaxial films with thicknesses ranging from 2-200 nm can be grown on SrTiO$_{3}$ substrates by pulsed-laser deposition.  Moreover, SRO has an unusually large Kerr coefficient that allows measurement of local changes in temperature of a few Kelvins in films that are only 2 nm thick \cite{magneto_optic}. Finally, as we discuss further below, the relevance of the magnetic DOF in thermal transport is accentuated in SRO compared with the more widely studied 3d ferromagnets such as Fe, Ni, and Co.

Fig. 1a shows the TRMOKE signal observed in an SRO film of thickness 200 nm, at several temperatures, $T$, below the Curie temperature of 150 K.  The curves show $\Delta \Theta_K$, as function of time $t$, after absorption of the pump pulse. $\Delta \Theta_K(t)$ is proportional to the $z$-component of \textbf{M} because the probe beam is at near normal incidence. The 200 GHz oscillations observed when $T<80$ K and $t<10$ ps correspond to damped precession of the magnetization, $\textbf{M}$ \cite{langner}. These oscillations are stimulated by a sudden change in the direction of the magnetic easy-axis, caused by absorption of the pump beam.  The fact that $\Delta \Theta_K (t)$ approaches a nonzero value when the oscillations have died down indicates that \textbf{M} has spiralled into alignment with a new local easy-axis direction. Below we show that the value that $\Delta \Theta_K$ reaches when the oscillations have died away provides a calibrated measure of the increase in local $T$ induced by absorption of the pump pulse.

In Fig. 1b we plot $\Delta \Theta_K$ at a fixed time delay of 25 ps as a function of $T$, for three different values of the pump laser intensity. (Shown in the inset to Fig. 1b is the $T$ dependence of the equilibrium Kerr rotation, $\Theta_K$, for the same 200 nm SRO sample). The peak in $\Delta \Theta_K(T)$ near the Curie temperature has been reported previously in a survey of TRMOKE in a variety of perovskite and spinel ferromagnets \cite{ogasawara}. In this work we focus on a new feature, which is the increase of $\Delta \Theta_K$ observed as $T\rightarrow 0$.

If $\Delta \Theta_K$ for $t>10$ ps is caused by local heating, we can use $\Theta_K(T)$ to calibrate the conversion from the TRMOKE signal to the local temperature, $T_l(t)$.  To test whether a local heating mechanism is consistent with the TRMOKE data, we consider the expected scaling of $\Delta \Theta_K(t)$ with bath temperature, $T$, and energy deposited by the laser pulse, $\Phi$. After $\Phi$ has been converted to heat, $T_l$ is determined by the relation, $\Phi = U(T_l) - U(T)$, where $U(T)$ is the internal energy. To exhibit the scaling relation, we express $T_l$ in terms of the inverse of $U(T)$,

\begin{figure}
\label{fig:fig1}
\includegraphics[width=2.5in]{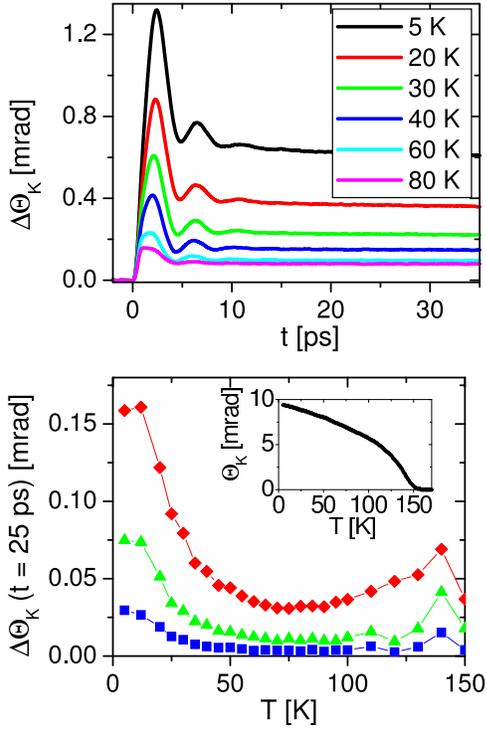}
\caption{(a) Thermally-induced change in $\Theta_{K}$ for different values of pump pulse intensity.  Inset: $\Theta_{K} (T)$ (b) Measured change in temperature in response to heating pulse.  Black lines represent a fit to $T^{3/2}$}
\end{figure}

\begin{equation}
\label{eq:TfU}
T_{l}(\Phi, T) = U^{-1}(\Phi + U(T)).
\end{equation}According to (\ref{eq:TfU}), plots of $T_l(\Phi,T)$ at fixed $T$ will fall on a single curve, $U^{-1}(\Phi)$, if shifted along the abscissa by $U(T)$.

\begin{figure}
\label{fig:fig2}
\includegraphics[width=2.5in]{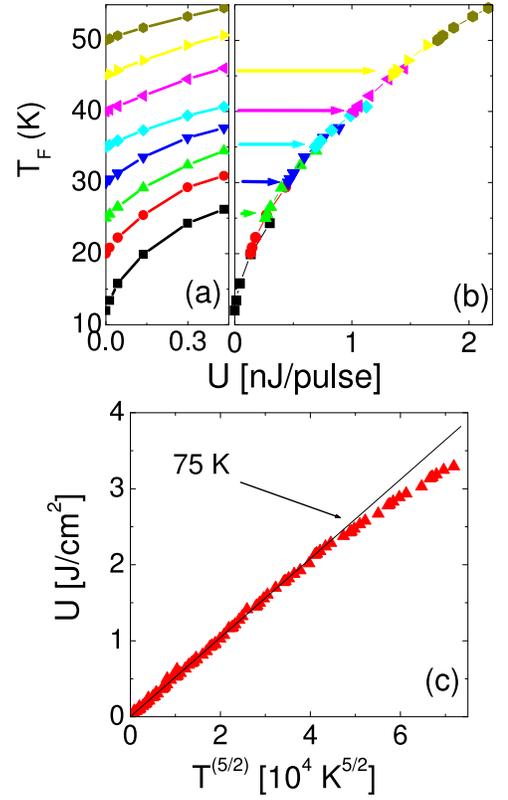}
\caption{(a) Unshifted measurements of final temperature as a function of pump laser intensity.  Lines represent different cryostat temperatures.  (b)  Data from (a) shifted along the $U$-axis to lie along a common curve. (c) Internal energy vs. T.  Data show a $T^{5/2}$ dependence below 75 K}
\end{figure}

In Fig. 2a (left panel), we plot $T_l(\Phi,T)$ at a delay of 25 ps as a function of $\Phi$ for different values $T$. The values of $T_l$ were determined by comparing $\Delta \Theta_K(t=25 ps)$, with the calibration data, $\Theta_K(T)$.  The result of shifting the curves to achieve the best collapse to a single curve is shown in the right panel of Fig. 2a. The data are seen to exhibit the scaling property expected for the local heating model.  A powerful feature of this scaling approach is that we can determine the internal energy function $U(T)$ from the magnitudes of the shifts required for the data collapse (shown as horizontal arrows in Fig. 2a). The internal energy as a function of $T$ determined by this procedure is shown in Fig. 2b.  When plotted as a function of $T^{5/2}$ the internal energy appears as a straight line with zero intercept for $T<75$ K.

The observation that $U(T)$ is proportional to $T^{5/2}$ (corresponding to specific heat $C\propto T^{3/2}$) indicates that in SRO thermal energy is stored predominantly in magnons for $T<75$ K. This contrasts with most ferromagnets, in which $C_p >> C_m$, where $C_p$ and $C_m$ are the phonon and magnon contributions, respectively. However, it is a natural consequence of the reversal in the ordering of the Curie and Debye temperature ($T_{\Theta}$) in the two classes of materials, that is $T_c<T_{\Theta}$ in SRO whereas $T_c\gg T_{\Theta}$ in the elemental 3d-transition metal ferromagnets.

From the preceding analysis, we conclude that the measurement of $\Delta\Theta_K(T,t)$ provides a direct probe of $T_l(T,t)$ in SRO thin films. In the following we explore the dynamics of $T_l$ as a function of time delay, substrate temperature, and film thickness.  In Fig. 3a we plot $\Delta \Theta_K(t)$ for 10 ps $<t<$500 ps, measured at 10 K, for films of different thickness in the range from 10-200 nm. The laser power is lowered so that $\Delta \Theta_K$ is directly proportional to $\Delta T_l(t)$, the difference between the instantaneous local temperature at the substrate temperature.  We find that $\Delta T_l(t)$ decays exponentially, with a time constant, $\tau$, that increases with sample thickness $L$.  Fig. 3b is a plot of $\tau$ as a function of $L$, showing that for $L>25$ nm, $\tau\simeq \alpha L+70$ ps, where $\alpha\simeq$ 2.7 ps/nm.  Exponential decay with a $\tau$ that increases linearly with thickness indicates that the interface between the substrate and the SRO acts as bottleneck that limits heat flow out of the film.  In this case the decay of $\Delta T_l$ is exponential with $\tau=RCL$, where $R$ is the thermal boundary resistance (TBR).

In Fig. 3c we plot $\tau/L$, (equal to the $RC$ product) as a function of temperature for films of thickness 5, 50, and 200 nm.  For the thickest film, the $RC$ product is almost independent of $T$ from 10-75 K, indicating that $R \propto 1/C$ over a broad range range of temperature.  For the thinner samples $RC$ increases with decreasing $T$ below about 50 K.

\begin{figure}
\label{fig:fig3}
\includegraphics[width=3.5in]{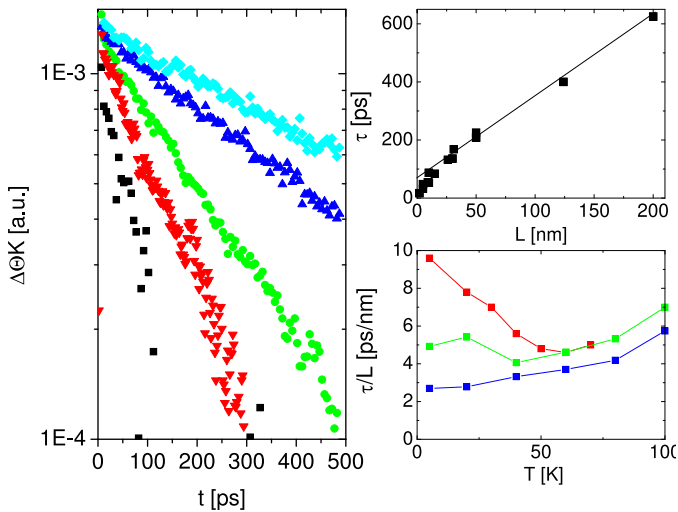}
\caption{(a) Semi-log plot of $\Delta \Theta_{K}(t)$ for 10, 30, 50, 125, and 200 nm samples (b)  Decay times vs. film thickness at 5 K (c) Decay times vs. temperature for 5, 50, and 200 nm samples}
\end{figure}

The TBR phenomenon is well-known for dielectric and metallic films on insulating substrates. In these systems TBR is understood to result from the scattering or specular reflection of phonons incident on the interface \cite{nanoscalereview, tbrreview}. This mechanism implies that $R \propto (|t|^2 C_p)^{-1}$, $t$ is the transmission coefficient for phonons at the interface.  For most materials the total specific heat $C$ is dominated by phonons, in which case $C_p$ cancels from the $RC$ product.  Thus $\tau$ essentially independent of $T$ is predicted, in agreement with many experimental observations.

As we have shown above, SRO exemplifies a system in which the specific heat is dominated by magnons rather than phonons. In this limit, $\tau(T)$ should vary as $C_m(T)/C_p(T)$, which is approximately proportional to $T^{-3/2}$.  Instead, as seen in Fig. 3c, $\tau$ is essentially independent of $T$, except for the thinnest film.

The lack of dependence of $\tau$ on $T$ suggests that the boundary resistance is not simply proportional to $C_p^{-1}$, but is somehow related to $C_m(T)$. At first glance, this seems unlikely, because magnons are confined to the magnetic film and cannot penetrate into the STO substrate.  This argument fails to consider the crucial fact that for heat to leave the SRO film, the energy stored in the magnon DOF must flow to the phonon DOF. For steady-state heat flows, it has been suggested \cite{majumdarreddy, ju} that disequilibrium between phonons and a confined DOF creates a mechanism for TBR different from the conventional phonon impedance mismatch picture. SRO/STO is an ideal testing ground this new mechanism because the fraction of thermal energy stored in the confined DOF is far larger than in the case of the metal film/dielectric interface.

Analysis of the flow of energy among electrons, phonons, and magnons requires a three-temperature model (3TM) \cite{three_temp}.  The 3TM assumes that each DOF is internally equilibrated so a temperature can be associated with each "fluid." We assume one-dimensional heat diffusion because the diameter of the excited region on the sample is $\sim$ 50 $\mu m$, much larger than both the optical penetration depth (37 nm) and the film thickness. Within the 3TM, the differential equation governing the electron temperature $T_e$, for example, is:

\begin{equation}
\label{eq:3TM}
C_{e} \frac{\partial T_{e}}{\partial t} = \frac{\partial}{\partial z} \left( \kappa_{e} \frac{\partial T_{e}}{\partial z} \right) - G_{ep}\Delta T_{ep}- G_{em}\Delta T_{em}+S.
\end{equation} Here $z$ is the normal coordinate, $\kappa_e$ is the electron thermal conductivity, and $S(z,t)$ represents heat input from the laser. $G_{ep}$ ($G_{em}$) governs the rate of heat flow from the electron to the phonon (magnon) DOF and $\Delta T_{ep}$ ($\Delta T_{em}$) is the electron-phonon (magnon) temperature difference.  Permutation of the subscripts generates equations for the time-evolution of the phonon and spin temperatures, $T_p$ and $T_m$, respectively.

To develop a picture of the heat flow dynamics, we have used a Dufort-Frankel finite-difference scheme \cite{trtransport} to integrate the 3TM equations.  In the course of analyzing the 3TM equations, we have found that a 2TM model is sufficient to capture the essential physics.  The electron and magnon DOF can be lumped together as single confined DOF, with thermal conductivity $\kappa_c\equiv\kappa_m+\kappa_e$, specific heat $C_c\equiv C_m+C_e$, and confined fluid temperature $T_c$. We note for SRO in the range 10 K$<T<$70 K that $\kappa_c\simeq\kappa_e$ and $C_c\simeq C_m$. For all calculations we have assumed that phonon thermal conductivity in the substrate is much larger than in the film, which enforces $\Delta T_l\simeq 0$ at the film/substrate interface.  Furthermore, we assume no heat flow from the surface of the film to vacuum, corresponding the boundary condition that $\Delta T^{\prime\prime} (z)$ vanishes as $z \rightarrow 0$.

In Figure 4 we plot $T_c(z)$ and $T_p(z)$ at fixed time delay, illustrating the two dynamical regimes that we observe. In the limit that $C_c<<C_p$, the 2TM predicts that $T_c(z)=T_p(z)$ (blue line) throughout the film. The temperature of both fluids varies on a length scale determined by the film thickness.  In the opposite limit, $C_c>>C_p$, which we believe is applicable to SRO, $T_c(z)$ and $T_p(z)$ are equal and nearly constant except near the interface, where $T_c$ remains constant while $T_p$ approaches $T$. In both regimes, $\Delta T_l$ decays exponentially with time, although the scaling of $\tau$ with $L$ is different in two cases, as we explain more fully below.

To understand these simulations, we note that the characteristic time for the two-fluids to reach quasiequilibrium is given by $\tau_{eq}=C_{eff}/G$, where $C_{eff}\equiv C_cC_p/C$. On this time-scale, thermal energy stored in the phonons propagates a distance $L_p=\delta(C_c/C)^{1/2}$.  The length scale $\delta\equiv (\kappa_p/G)^{1/2}$ is the disequilibrium length identified previously \cite{majumdarreddy, ju}. When $C_c<<C_p$, we have $L_p<<\delta$, and the temperatures of the two-fluids are locked together throughout the film. In the opposite regime, where most of the energy is stored in the confined DOF, $L_p\rightarrow\delta$.

While our simulations indicate that the decay of $\Delta T_l$ with time is exponential in both regimes, we find different scaling with $L$; $\tau \propto L^2$ when the phonons dominate the specific heat, whereas $\tau \propto L$ when the magnons dominate. The contrasting spatial profile of $T_p$ is the key to the different scaling. When $C_c<<C_p$, $T_p(z)$, $\tau\approx L^2/D_p$, where $D$ is the thermal diffusivity. In the opposite limit the phonon heat current $J$ is approximately $\kappa_p T_p(0)/\delta$ (where $\kappa_p$ is the phonon thermal conductivity) within a length $\delta$ from the interface and is essentially zero elsewhere. The thermal time constant is governed by the relation $\tau=Q/\dot{Q}$, where $Q$ is thermal energy per unit area. Substituting $\dot{Q}=J$ and $Q=C_c L T_p(0)$ we obtain,

\begin{equation}
\label{eq:tauneq}
\tau = \frac{C_cL \delta}{\kappa_p} = \frac{C_c L}{\sqrt{G \kappa_{p}}}.
\end{equation}

\begin{figure}
\label{fig:fig4}
\includegraphics[width=2.5in]{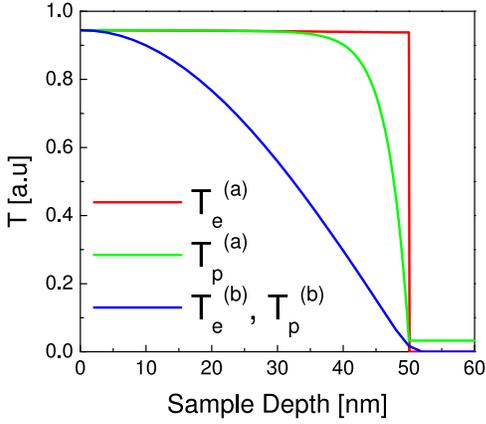}
\caption{Temperature profiles for confined fluid and phonons for (a) $C_c \ll C_p$ and $\kappa_{c} < \kappa_{p}$. (b) $C_c \gg C_p$ and $\kappa_{c} > \kappa{p}$.  The temperature profile near the surface ($z = 0$) is dominated by the phonons in (a), and by confined fluid in (b).  The shapes of these profiles remain constant throughout the decay.  In both cases, $\kappa_p$ in the substrate is greater than that in the film.}
\end{figure}

Eq. 4 is the key result that emerges from the 2TM model in the $C_c>>C_p$ regime. While the linear dependence of $\tau$ on $L$ indicates the existence of TBR, Eq. 4 is derived from a model that assumes perfect phonon transmission at the interface.  The \emph{effective} TBR seen here is the same as predicted to arise from disequilibrium between free and confined DOF's, for steady-state heat flow \cite{majumdarreddy, ju}. It is somewhat surprising to see the same effect in time-dependent heat flow, where it has been argued that two-fluid disequilibrium does not influence the rate of cooling when $t>\tau_{eq}$.  However, our analysis shows that the presence of the interface leads to persistence of disequilibrium on the much longer time scale, $\tau_{eq}(L/\delta)$.

Next, we consider the dependence of $\tau$ on $L$ and $T$ in the light of the discussion above. To compare the measured $\tau(L)$ with Eq. 4, we assume that the effective phonon mean-free path, $l_p$ is given approximately by the relation $l_p^{-1}=l_{p0}^{-1}+L^{-1}$, where $l_{p0}$ is the mean-free path in the absence of interfaces.  Substituting the effective thermal conductivity $\kappa_p l_p/l_{p0}$ into Eq. 4 yields $\tau(L)=\alpha (L+l_{p0})$ for $L>>l_{p0}$ and $\tau(L)=\alpha (l_{p0}L)^{1/2}$ for $L<<l_{p0}$, where $\alpha\equiv (C_c/G\kappa_p)^{1/2}$. The predicted linear to square root crossover in the dependence of $\tau$ on decreasing $L$ is fully consistent with the experimental data presented in Fig. 3b. We identify the crossover value of $L$ of approximately 15 nm with the phonon mean-free path at 5 K. To make further quantitative comparison with experiment, we note that from the measured value $\alpha$=2.7 ps/nm and literature value $C=10^3$ W/m-K, we obtain $\sqrt{G\kappa_p}=4\times 10^5$ W/m$^2$-K. To obtain $G$ we need $\kappa_p$, which is not easy to extract from experiments, as they measure the total $\kappa$. Nevertheless, we can base an estimate on the kinetic theory relation, $\kappa_p=C_pv_s l_p$; substituting $l_p=15$ nm, $v_s=10^3$ m/s, and $C_p=24$ W/m$^2$-K (based on $\Theta_D=$370 K \cite{allen}) yields $\kappa_p\approx$1.2$\times10^{-4}$ W/m-K.  This estimate, in turn, yields $G\approx1.6\times10^{15}$ W/m$^3$-K, which is consistent with equilibration rates seen in other magnetic materials \cite{ogasawara, bigot, beaurepaire}.

Finally, we return to our original motivation for considering disequilibrium effects, the weak temperature dependence of $\tau$ for thick films.  Above we have shown that in the thick film limit $\tau/L\approx (C_c/C_p^{1/2})(G v_s l_p)^{-1/2}$. According to this formula, there is an accidental cancelation of the strongly $T$-dependent terms; the $T^{3/2}$ dependence of the magnon specific heat cancels the square root of the $T^3$ dependence of the phonon specific heat.  The residual weak $T$ dependence of $\tau/L$ (see Fig. 3c) is associated with the term $(G v_s l_p)^{-1/2}$. For thin films, the increase in $l_{p0}(T)$ with decreasing $T$ is cutoff at $L$ and below this $T$, $\tau/L$ becomes simply proportional to $G^{-1/2}$.  Again referring to Fig. 3c, and focusing on the 5 nm sample, we associate the change in slope at 50K with the crossover between $l_{p0}>L$ and $l_{p0}<L$ regimes.

To summarize, we have used TRMOKE to time-resolve the local $T$ in thin films of the ferromagnet SRO. The scaling of $T_l$ with laser intensity and substrate $T$ shows that the specific heat of SRO is dominated by the magnon contribution in a broad $T$ range.  The flow of heat from film to substrate appears to be limited by an effective thermal boundary resistance arising from disequilibrium between confined and propagating degrees of freedom.

\begin{acknowledgments}
This research is supported by the US Department of Energy, Office of Science.
\end{acknowledgments}

\end{document}